\documentclass[prd,twocolumn,showpacs,nofootinbib]{revtex4}

\usepackage{amsfonts}
\usepackage{amsmath}
\usepackage{amssymb}
\usepackage{bm}
\usepackage{dcolumn}
\usepackage{epsfig}
\usepackage{graphicx}
\usepackage{graphics}
\usepackage[latin1]{inputenc}
\usepackage{latexsym}
\usepackage{rotating}
\usepackage{hyperref}
\usepackage[usenames]{color}

\newcommand\be{\begin{equation}}
\newcommand\ba{\begin{eqnarray}}
\newcommand\ee{\end{equation}}
\newcommand\ea{\end{eqnarray}}

\newcommand{\GR}{{\mbox{\tiny GR}}}
\newcommand{\ppe}{{\mbox{\tiny ppE}}}

\begin{document}
\title{Gravitational Wave Tests of Strong Field General Relativity with Binary Inspirals: \\
Realistic Injections and Optimal Model Selection}

\author{Laura Sampson}
\affiliation{Department of Physics, Montana State University, Bozeman, MT 59717, USA.}

\author{Neil Cornish}
\affiliation{Department of Physics, Montana State University, Bozeman, MT 59717, USA.}

\author{Nicol\'as Yunes}
\affiliation{Department of Physics, Montana State University, Bozeman, MT 59717, USA.}

\date{\today}

%%%%%%%%%%%%%%%%%%%%%%%%%%%%%%%%%%
%%%%%%%%%%%%%%%%%%%%%%%%%%%%%%%%%%
%%%%%%%%%%%%%%%%%%%%%%%%%%%%%%%%%%
\begin{abstract}

We study generic tests of strong-field General Relativity using gravitational waves emitted during the inspiral of compact binaries. 
Previous studies have considered simple extensions to the standard post-Newtonian waveforms that differ by a single term in the
phase.  Here we improve on these studies by (i) increasing the realism of injections and (ii) determining the optimal 
waveform families for detecting and characterizing such signals. 
We construct waveforms that deviate from those in General Relativity through a series
of post-Newtonian terms, and find that these higher-order terms can affect our ability to test General Relativity, in some cases by
making it easier to detect a deviation, and in some cases by making it more difficult. 
We find that simple single-phase post-Einsteinian waveforms are sufficient for detecting deviations from General Relativity, and there is little to be gained from using
more complicated models with multiple phase terms.
The results found here will help guide future attempts to test General Relativity with advanced ground-based detectors. 
  
\end{abstract}
%%%%%%%%%%%%%%%%%%%%%%%%%%%%%%%%%%
%%%%%%%%%%%%%%%%%%%%%%%%%%%%%%%%%%
%%%%%%%%%%%%%%%%%%%%%%%%%%%%%%%%%%

\pacs{04.80.Cc,04.80.Nn,04.30.-w,04.50.Kd}
\maketitle

%%%%%%%%%%%%%%%%%%%%%%%%%%%%%%%%%%
%%%%%%%%%%%%%%%%%%%%%%%%%%%%%%%%%%
%%%%%%%%%%%%%%%%%%%%%%%%%%%%%%%%%%
\section{Introduction}

Einstein's General theory of Relativity (GR) has weathered an array of increasingly stringent tests since the theory first gained prominence in November 1919,
when reports of Eddington's expedition appeared in newspapers around the world: ``Revolution in science -- New theory of the Universe -- Newtonian ideas overthrown''. Subsequent observations
have continued to strengthen the case for Einstein's theory, though observations have yet to probe the dynamical, non-linear regime where
the most revolutionary aspects of the theory take hold. For example, GR has passed all Solar System tests with flying colors, but these are based on stationary, weak, and linear gravitational fields, where characteristic velocities are small relative to the speed of light~\cite{lrr-2006-3}. The theory has also passed all binary pulsar tests, but these systems have gravitational fields that are quasi-stationary and only moderately-strong, with characteristic velocities of $\sim 0.1\%$ the speed of light~\cite{lrr-2006-3,Kramer:2006nb}. In the near future, gravitational wave (GW) observations will test GR in a regime that has so-far evaded observation: the {\emph{strong-field}}, where the gravitational field is of order unity and velocities approach the speed of light.

Compact binary coalescences, the slow inspiral and merger of black holes (BHs) and/or neutron stars (NSs), will be strong sources of GWs, and these will be excellent tools for testing GR. During the inspiral phase, the binary components have orbital velocities ranging from $1\%$ to $\sim 50\%$ the speed of light, which leads to strong and dynamically evolving gravitational fields. These GW signals evolve through thousands of radians of phase in the most sensitive band of ground-based detectors, such as aLIGO and aVIRGO, with signal-to-noise ratios (SNRs) that will allow us to extract signal parameters with good accuracy. Thus, even small differences in the dynamics of the gravitational theory can lead to large accumulated effects in the waveform during the inspiral.

Despite their promise, GW tests of GR are, unfortunately, very difficult to carry out, for two main reasons. One reason is purely theoretical - we currently lack candidate alternative theories that are particularly appealing. Instead, we have many models that are either heavily constrained, like scalar-tensor theories~\cite{lrr-2006-3}, or that have theoretical issues, such as knowledge only of their effective, low-curvature form~\cite{Alexander:2009tp}. The other cause of difficulty lies in the data analysis. Most techniques for detecting and characterizing GW observations require accurate templates to identify weak signals buried in the instrument noise. Given the already large parameter dimensionality of the GR waveform models, and the wide variety of modified gravity theories~\cite{Will:1994fb, Scharre:2001hn, Will:2004xi, Berti:2005qd, Yagi:2009zm, Will:1997bb, Stavridis:2009mb, Arun:2009pq, Keppel:2010qu, Alexander:2007:gwp,Yunes:2010yf,Yunes:2009hc,Sopuerta:2009iy,Alexander:2009tp,Yagi:2012vf,Yunes:2009bv,Bekenstein:2004ne}, the construction of individual template banks for all possible non-GR models is simply not feasible.      

A much more appealing alternative is to devise a generic non-GR template family with which to model the signals,  and allow the data
to select the appropriate model via Bayesian inference. The first such model was proposed by Arun {\emph{et al}}~\cite{Arun:2006hn,Arun:2006yw,Mishra:2010tp}, where the coefficients in the post-Newtonian (PN) expansion of the phase were independently fitted for. However, the structure of the PN series does not allow for all known modified gravity deviations, including potentially interesting ones such as the emission of dipolar radiation predicted in scalar-tensor theories. For this reason,  Yunes and Pretorius~\cite{Yunes:2009ke} developed the so-called parameterized post-Einsteinian (ppE) framework, which allows for a wide range of deformations to the amplitude and phase of the waveform. In the inspiral phase, these can be represented through a polynomial in the GW frequency, with free constants, or ppE parameters, that represent the amplitude and the frequency exponent of the deformations~\cite{Yunes:2009ke}. The simplest ppE inspiral waveform in the Fourier domain has the form
\be
\tilde{h}^{\ppe} = \tilde{h}^{\GR} \left(1 + \alpha u^{a}\right) e^{i \beta u^{b}}\,,
\label{eq:simple-ppE}
\ee
where $u = (\pi {\cal{M}} f)^{1/3}$ is a dimensionless velocity, ${\cal{M}} = \eta^{2/5} M$ is the chirp mass, $\eta = m_{1} m_{2}/(m_{1}+m_{2})^{2}$ is the symmetric mass ratio, $m_{1,2}$ are the component masses, and $f$ the GW frequency. The Fourier transform of the GR waveform is here $\tilde{h}^{\GR}$, while $(\alpha,a,\beta,b)$ are ppE parameters. Clearly, in the limit $(\alpha,\beta) = (0,0)$, one recovers the GR prediction, while for other values of the ppE parameters one recovers the leading-order waveforms of all known modified gravity theories.

The first data analysis implementation of the ppE framework was carried out by Cornish, {\emph{et al}}~\cite{Cornish:2011ys}, where ppE waveforms of the form of Eq.~\eqref{eq:simple-ppE} were used both in the generation of the simulated signals, and in the extraction of the model parameters in a Bayesian model
selection framework. This study was a proof-of-principle that the ppE framework can be successfully implemented to carry out tests of GR. A second study shortly followed~\cite{Li:2011cg} that confirmed the results of Cornish, {\emph{et al}} and extended them to include lower SNR signals and multiple detections. While this study also used the simple one-phase ppE model of \eqref{eq:simple-ppE} for the signal injections, the models used to analyze the simulated data included more complicated ppE waveform models with multiple phase corrections.

In this paper we revisit the ppE framework and carry out a more realistic data analysis study. First, we examine the effect of more realistic non-GR injections that include modifications to several terms in the PN GR phase, instead of a single one. Generic deviations from GR will be characterized by an infinite number of phase corrections. Ground-based detectors will not be sensitive to all of them, just as they are not sensitive to GR signals to arbitrarily high PN order. The presence of the first few higher-order terms can affect our ability to test GR. We find that the presence of multiple phase modifications will improve our chances of detecting departures from GR if they are of the same sign. However, if the phase modifications are of alternating sign, they can cancel out to some degree, and make a non-GR signal appear to be well described by GR. 

As something of an aside, we consider the issue of adding explicit noise realizations to the simulated signals, especially for low SNR signals. This is done because some concerns
have been voiced about the conclusion of the Cornish {\emph{et al}}~\cite{Cornish:2011ys} work due to the relatively high SNR of the signals used, and
their technique of accounting for the noise solely through the weighting of the likelihood function. We analytically and numerically show that the conclusions of~\cite{Cornish:2011ys} remain unaffected when adding an explicit noise realization. We also show that these results scale linearly with SNR down to values close to the detection threshold.

We then tackle the problem of determining the {\emph{optimal}} ppE model for detecting departures from GR.  On the one hand, including additional phase terms
will improve the fit and increase the likelihood. On the other hand, adding additional parameters to the model incurs an ``Occam penalty''. We find, on balance,
that in almost all cases, templates with only one ppE parameter are preferred over those with multiple parameters. These suggests that the simple one-parameter ppE model may well be the ideal one to search for GR deviations in early data from advanced detectors. 

The remainder of this paper is organized as follows.  
Section~\ref{realistic} builds non-GR injections and studies their effect on signal extraction and the detection of departures from GR.
Section~\ref{sec:noise-SNR} considers the effects of adding explicit noise realizations to the signals, and how the strength of the signal affects our ability to test GR. 
Section~\ref{sec:optimal-fitting} studies different ppE waveform models to determine the optimal one for performing GW tests of GR.
Section~\ref{sec:conclusions} concludes and points to future directions for research.
Throughout this paper we use geometric units with $G=c=1$. 

%%%%%%%%%%%%%%%%%%%%%%%%%%%%%%%%%%
\section{Realistic Signal Injections}\label{realistic}

The simplest ppE waveform family presented in the introduction is not sufficiently complex to represent a realistic alternative gravity theory. This is because modified gravity theories will differ from GR by an infinite series of terms in both the amplitude and the phase. We expect that an alternative theory of gravity will give rise to waveforms where the amplitude and phase depend on one or more fundamental coupling constants multiplied by functions of the system parameters.  Thus, if one wishes to use a ppE-type template to inject non-GR signals, one must consider more complex ppE models, such as Eq.~$(46)$ in~\cite{Yunes:2009ke}, namely Eq.~\eqref{eq:simple-ppE} with the replacements~\cite{Yunes:2009ke}
\be
\alpha u^{a} \to \sum_{i=0}^{N} \alpha_i u^{a_{i}}\,, \quad \beta u^{b} \to \sum_{i=0}^{N} \beta_i u^{b_{i}}\,,
\label{eq:gen-ppE}
\ee
where the $\alpha, \beta$'s depend on a universal coupling constant $\kappa$, and functions of the system parameters $\vec\lambda$:
\ba
&& \alpha_i(\kappa, \vec{\lambda})  = \kappa \sum \phi_i(\vec{\lambda})\, \nonumber \\
&& \beta_i(\kappa, \vec{\lambda})  = \kappa \sum \theta_i(\vec{\lambda})\, .
\ea
The functions $\phi_i(\vec{\lambda}), \theta_i(\vec{\lambda})$ can be computed for specific theories, but their general form is unknown.
So while $\kappa$ takes a single value for a particular theory, the $(\alpha_i, \beta_i)$ constants will vary from detection to detection depending on
the masses, spins and other parameters that describe the system. In some theories there will be more than one additional coupling constant $\kappa$, 
but here we will assume that one sector of the modified theory dominates and consider only a single series of correction terms.
With a large number of high SNR detections, it may be possible to infer the functional form of 
$(\phi_i(\vec{\lambda}), \theta_i(\vec{\lambda}))$. However, since our immediate
concern is in deciding if the data is consistent with the prediction of GR, we will argue that it is best to use a much simpler waveform for the initial tests.

The ppE exponents $(a_i, b_{i})$ are real numbers that give the effective PN order at which the non-GR modification enters the signal, while the ppE
amplitude parameters $(\alpha_i, \beta_{i})$ are real numbers that indicate the strength of the modification, in turn controlled by the overall coupling strength $\kappa$. In principle, we could extend the sum in Eq.~\eqref{eq:gen-ppE} to infinity, but in practice, realistic detectors are sensitive only to a finite number of terms in the
phase and amplitude. The injected signals then consist of a GR waveform with its amplitude and phase modified by a series of ppE corrections.

Several simplifications can be made to the general waveform presented above. First, for quasi-circular inspiral signals, Chatziioannou, {\emph{et al}}~\cite{Chatziioannou:2012rf} have argued that analyticity demands that the exponents $(a_i, b_{i})$ take on integer values with possible logarithmic corrections
 (just as the PN expansion in GR comes in integer powers of $u$ and products of integer powers of $u$ with $\log u$, where recall here that $u$ is related to the orbital velocity). Second, ground-based advanced detectors will be of limited sensitivity, rarely being sensitive to more than the first three terms in the PN expansion, and usually being much more sensitive to the phase evolution than they are to the amplitude evolution. Thus, we choose to simplify the analyses by truncating the sum at three terms and setting $\alpha_{i} = 0$. The injections are then given by Eq.~\eqref{eq:simple-ppE} but with the replacement 
\begin{align}
\beta u^{b} &\to \sum_{i=0}^{2} \beta_i u^{b+i} = \beta_{b} u^{b} + \beta_{b+1} u^{b+1} + \beta_{b+2} u^{b+2}\,,
\label{eq:phase-cor}
\end{align}
and $\alpha  = 0$, where in the last equality the Einstein summation convention is not assumed. Written in this way, $\beta_{b}$ is always proportional to $u^{b}$ for any $b$.  Previous investigations have been restricted to signals with only one ppE correction injected, which reduces to Eq.~\eqref{eq:phase-cor} when one retains only the first term in the sum. As argued above, this is far from realistic for a modified gravity injection and we will show that the higher-order terms can have a significant effect on the analysis.

Ultimately the claim that a detection is in agreement (or conflict) with GR comes down to model selection. Does GR describe the data best or does another model do a better job? In Bayesian statistics~\cite{cornish:083006,Cornish:2011ys,2011arXiv1101.1391D}, model selection is performed via the calculation of the Bayes factor, which is simply the ``betting odds" of one model against another. For instance, if the Bayes factor between GR and a non-GR model is 100, and you originally gave both possibilities equal odds, then there is a 100:1 odds ratio that GR better describes the data than the other model. In this case, you would be well-advised to put your money on GR. There is no prescription for deciding what Bayes factor is required before we should consider one model ``right'' and another ``wrong''. However, in the case of a well-tested theory like GR being brought into question by, for instance, a GW signal, it is likely that the scientific community would require a detection that gives us a fairly high Bayes factor in favor of the non-GR model to overcome the prior belief in GR being the correct theory. In order to determine whether more ppE terms in an injection affect the detectability of a deviation from GR, we need to see how these different types of injections affect the Bayes factor. 

Throughout this paper, Bayes factors are calculated using the Savage-Dicke density ratio ~ \cite{Dickey, cornish:083006} and/or Reversible Jump Markov Chain Monte Carlo (RJMCMC) \cite{Sambridge, cornish:083006}. In the Savage-Dicke method, the Bayes factor between two nested models, i.e. model X and model Y that differ only by the addition of a parameter to model Y, is calculated by comparing the weight of the marginalized posterior to the weight of the prior distribution for the ``extra'' parameter at the value that this parameter takes on for the lower-dimensional model:
\be
B_{XY} \approx \frac{p(\kappa=0|s)}{p(\kappa=0)}
\ee
In our case, the extra parameter is the coupling strength $\kappa$, which has the GR limit $\kappa = 0$, and hence $\beta_i =0$.  To calculate the Bayes factors this way, we run a MCMC search using ppE templates in order to generate the posterior distribution for $\beta_i$, and then calculate the posterior weight in this distribution at $\beta_i = 0$. We then compare this posterior weight to the prior density at this point. We here use a flat prior distribution between $-5.0$ and $5.0$ for all $\beta$ values. The main advantage to this method over other possibilities is that it only requires exploration of the higher-dimensional space.  

All tests in this section use GWs emitted by a NS-NS binary with $\approx1.4 M_{\odot}$ component masses in the inspiral phase with SNR $\sim 12$. We model all waveforms with a quadrupolar, adiabatically quasi-circular waveform, with a 3.5PN-accurate phasing, but neglecting PN amplitude correction and spin effects, and truncating all evolution at the Schwarzschild test-particle innermost stable circular orbit. The waveforms are then described by nine source-parameters: the chirp and the reduced mass; the time and phase of coalescence; two sky-position angles; the inclination angle and the GW polarization angle; and the luminosity distance (see~\cite{Cornish:2011ys} for a similar waveform prescription). In addition to these we have the ppE phase parameters of Eq. (\ref{eq:phase-cor}). We consider a three detector network of second-generation detectors, such as aLIGO at Hanford, aLIGO at Livingston, and aVirgo, with identical broadband-configuration spectral densities, as in our previous paper \cite{Cornish:2011ys}, assuming the noise to be Gaussian and stationary.  Table~\ref{table:SP0} shows the system parameters for all systems studied in this paper (masses are listed in solar masses, and luminosity distances are in megaParsecs).

In this paper, we examine two factors that influence the outcome - the signs of the different phase corrections, and their relative magnitude. We begin by exploring the effects of injecting phase corrections with the same or differing signs. In particular, let us study the effect that this relative sign has on the detectability of a non-GR behavior. We will then explore the difference between non-GR phase corrections that either shrink in magnitude at higher PN order, stay at approximately the same magnitude, or grow in magnitude at higher PN order. 

\begin{table*}[ht]
\centering  % used for centering table
\begin{tabular}{l c c c c c c c c c c c c} % centered columns (4 columns)
\hline\hline                        %inserts double horizontal lines
Signal & $\alpha$ & $\phi_L$ & $\phi_c$ & $m_1 (M_{\bigodot})$ & $m_2 (M_{\bigodot})$ & $\log(D_L) ({\rm{Mpc}})$ & $t_c$ & $\delta$ &$\theta_L$ & $\beta_{-3}$ & $\beta_{-2}$ & $\beta_{-1}$ \\ [0.5ex] % inserts table 
%heading
\hline                  % inserts single horizontal line
One ppE Term & 1.0   & 4.76 & 1.9 &1.62 &1.73 & 3.96 &5.58 &0.77 &-0.43 &0.01  &0.0 &0 \\ 
Alternating Sign        & 1.0   & 4.76 & 1.9 &1.62 &1.73 & 3.96 &5.58 &0.77 &-0.43 &0.01 &-0.04    &0  \\
Same Sign  & 1.0   & 4.76 & 1.9 &1.62 &1.73 & 3.96 &5.58 &0.77 &-0.43 &0.01  &0.04   &0  \\ [1ex]      % [1ex] adds vertical space
\hline %inserts single line
\hline %inserts single line
Convergent & 1.0   & 4.76 & 1.9 &1.62 &1.73 & 3.96 &5.58 &0.77 &-0.43 &0.01  &0.005 &0 \\ 
Critical          & 1.0   & 4.76 & 1.9 &1.62 &1.73 & 3.96 &5.58 &0.77 &-0.43 &0.01 &0.08    &0  \\
Asymptotic   & 1.0   & 4.76 & 1.9 &1.62 &1.73 & 3.96 &5.58 &0.77 &-0.43 &0.01  &0.25   &0  \\ [1ex]      % [1ex] adds vertical space
\hline %inserts single line
\hline                  % inserts single horizontal line
GR Source & 3.95 & 4.14 & 0.68 &1.45 &1.43 &0.9 &3.41 &-0.66 &0.76 &0  &0 &0 \\ % inserting body of the table
\hline %inserts single line
\end{tabular}
 % is used to refer this table in the text
\caption{\label{table:SP0} Source parameters for sources used in Fig.~\ref{BFthreeinj} (top), Fig.~\ref{pp_pm_mm} (middle) and Figs.~\ref{noise_BF}, \ref{noise_3seeds}, and \ref{beta1Mcorr} (bottom).}
\end{table*}

We begin by examine how the relative sign of the phase corrections affects the detectability of departures from GR. To do this, we consider three non-GR injections: 
\begin{itemize}
\item {\bf{Case i.}} A ppE waveform with a single ppE phase term ($b=-3$), with magnitude controlled by $\beta_{-3}$.
\item {\bf{Case ii.}} A ppE waveform with two ppE phase terms ($b = -3$ and $b = -2$), with $\beta_{-3}$ and $\beta_{-2}$ of the same sign.
\item {\bf{Case iii.}} A ppE waveform with two ppE phase terms ($b = 3$ and $b = +2$), with $\beta_{-3}$ and $\beta_{-2}$ of different sign.
\end{itemize} 
We choose these values of $b$ because, for $b < -5$, $\beta_b$ is already well-constrained by binary pulsar observations, as demonstrated in~\cite{2010PhRvD82h2002Y,Cornish:2011ys}. Case (i) is the type of injection that has been explored in previous work. Cases (ii) and (iii) include higher-order phase corrections, but differ in their relative sign.

\begin{figure}[ht]
\begin{center}
\begin{tabular}{cc}
\epsfig{file=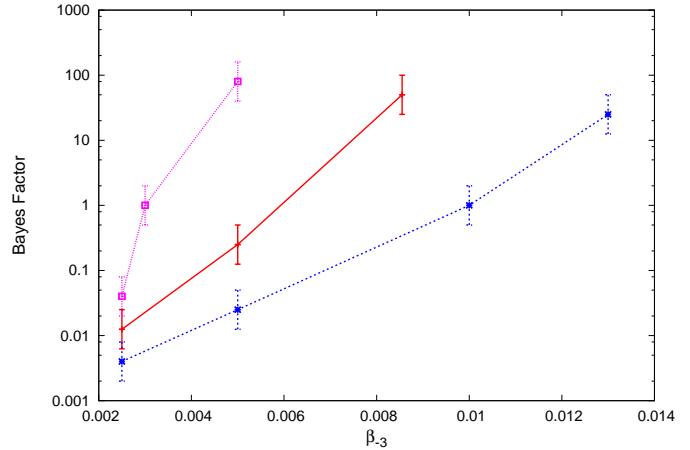,angle=-90, width=0.5\textwidth} 
\end{tabular}
\end{center}
\vspace*{-0.2in}
\caption{\label{BFthreeinj} (Color Online) Bayes factors between a GR model and a one-parameter ppE model for three different ppE signal injections. The dotted (magenta) line corresponds to an injection with the two positive ppE terms $\beta_{-3} >0$ and $\beta_{-2} > 0$ (case ii), the solid (red) line corresponds to the single, positive ppE term $\beta_{-3}>0$ (case i), and the dashed (blue)  line corresponds to the two ppE terms of alternating sign $\beta_{-3} > 0$ and $\beta_{-2} <0$ (case iii). System parameters for the systems studied here are listed in Table \ref{table:SP0}. As expected, the signal with ppE terms of alternating sign is harder to distinguish from GR, as evidenced by its Bayes factor growing the slowest with the magnitude of $\beta_{-3}$. }
\end{figure}

Figure~\ref{BFthreeinj} shows the Bayes factors between GR and a one-parameter ppE template family with $b = -3$ and ppE parameter $\beta_{-3}$ for the three injections discussed above. The error bars in this figure are estimated by calculating the Bayes factors using multiple MCMC runs with different random seeds. The spread in the calculated values are reflected in the error bars.  Observe that when the injection contains ppE corrections of the same sign (dotted, magenta curve), these add up to make the signal more discernible from GR. In this case, the Bayes factor crosses $10$ for the smallest value of $\beta_{-3}$. Therefore, if $(\beta_{b},\beta_{b+1})$ share the same sign, we can detect deviations from GR with lower strengths than if there were only one phase correction. On the other hand, observe how when the non-GR signal contains alternating sign GR modifications (dashed blue line), these have the effect of partially canceling the non-GR effect out. In this case, the Bayes factor crosses $10$ for a much larger value of $\beta_{-3}$. Therefore, if the corrections have alternating signs, e.g.~if $(\beta_{b},\beta_{b+1})$ have different signs, then our ability to detect departures from GR is reduced. The sign of the ppE amplitude exponent also affects the PDFs of the recovered $\beta_{i}$ parameters, as we will see below.

The relative magnitudes of the terms also affects the analysis. Concentrating on the multi-term ppE models of Eq. (\ref{eq:phase-cor}), we
define three cases, depending on the relative magnitude of these exponents in the series expansion: 
\begin{itemize}
\item {\bf{Convergent Case}}: Injections where the ppE terms get smaller as the PN order increases, i.e.~$\beta_{b} > \beta_{b+1} > \beta_{b+2}$.
\item {\bf{Critical Case}}: Injections where the ppE terms remain of about the same size as the PN order increases, i.e.~$\beta_{b} \sim \beta_{b+1} \sim \beta_{b+2}$.
\item {\bf{Asymptotic Case}}: Injections where the ppE terms get bigger as the PN order increases, i.e.~$\beta_{b} < \beta_{b+1} < \beta_{b+2}$.
\end{itemize}
Obviously, there are an infinite number of ways to choose how large the $\beta_{i}$ constants are relative to each other, but the classification defined above provides a useful summary. More concretely, we here define {\emph{convergent}}\footnote{We use the words ``convergent,'' ``critical'' and ``asymptotic'' loosely here. These names do not necessarily imply that the series posses these properties.} cases as those where the ppE terms injected have $\beta_{n+1} < (u_{\max})^{-b_{n}}$, where $u_{\max} = \pi \mathcal{M} f_{\max}$. Similarly, {\emph{critical}} cases are defined such that $\beta_{n+1} \approx (u_{\max})^{-b_{n}}$, while {\emph{asymptotic}} cases have $\beta_{n+1} < (u_{\max})^{-b_{n}}$.

An alternative and roughly equivalent way to define these three different cases is by the number of {\emph{useful cycles}} of phase \cite{2000PhRvD..62h4036D} that accumulate during the signal for each correction to the phase. The number of useful cycles is defined via 
\be
N_{useful} \equiv \left( \int_{F_{\min}}^{F_{\max}} \frac{df}{f} \frac{a^2(f)}{S_n(f)} \frac{d\phi}{2\pi df} \right) \left( \int_{F_{\min}}^{F_{\max}} \frac{df}{f} \frac{a^2(f)}{f S_n(f)} \right)^{-1}
\ee
where $|\tilde{h}(f)|^2 = N(f) a^2(f)/ f^2$ is the squared modulus of the frequency domain GW signal, and $N(f) = (f/2\pi)(d\phi/df)$.
This quantity tells us about the phase accumulated from each PN (or ppE) term during the course of the signal, weighted by the sensitivity of the detector to different parts of frequency space. Tables of the number of useful cycles of phase for each system analyzed in this paper are included in this section. ``Convergent" signals are those for which the number of useful cycles due to the non-GR phase corrections decreases at higher order. ``Critical" signals have roughly the same number of useful cycles at each order. ``Asymptotic" signals have larger numbers of useful cycles from the non-GR phase at higher orders. 

\begin{table}[ht]
%\centering  % used for centering table
\begin{tabular}{l c c c} % centered columns (4 columns)
\hline\hline                        %inserts double horizontal lines
Signal & $\phi_{-3}$ & $\phi_{-2}$& $\phi_{-1}$ \\ [0.5ex] % inserts table 
%heading
\hline                  % inserts single horizontal line
Convergant & 0.312 & 0.012 & 0  \\ % inserting body of the table
Critical          & 0.312 & 0.194 & 0  \\
Asymptotic   & 0.312 & 0.607 & 0 \\ [1ex]      % [1ex] adds vertical space
\hline %inserts single line
\end{tabular}
\label{table:usecyc12} % is used to refer this table in the text
\caption{Number of useful cycles from the different injected ppE terms - Fig \ref{BFthreeinj} and Fig \ref{pp_pm_mm}.}
\end{table}

\begin{figure}[htb]
\begin{center}
\begin{tabular}{cc}
\epsfig{file=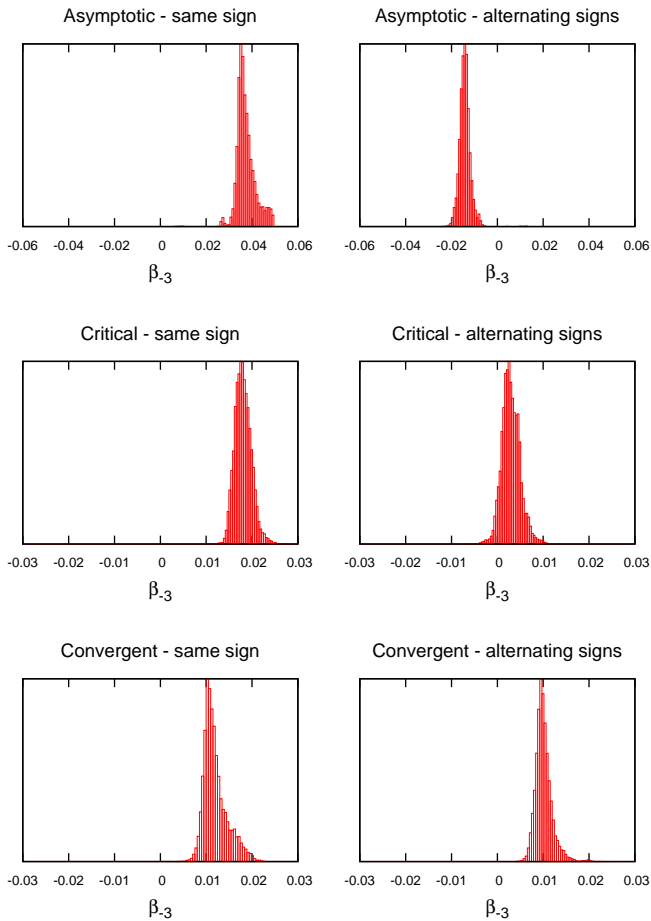, width=0.5\textwidth,angle=0} 
\end{tabular}
\end{center}
\vspace*{-0.2in}
\caption{ \label{pp_pm_mm}  The PDF's for $\beta_{-3}$ in a one-parameter ppE template recovered from MCMC searches on injections containing two ppE parameters ($b=-3$ and $b=-2$). The plots on the left are for injections containing two ppE parameters of the same sign, and on the right of opposite signs. The more weight in the PDF at $\beta = 0$, the lower the Bayes factor in favor of a non-GR signal. In the critical case, we find that alternating signs in the phase corrections can cause a non-GR signal to be indistinguishable from a GR one. In the convergent and asymptotic cases, this does not occur. System parameters for this figure are the same as in Figure \ref{BFthreeinj}, also listed in Table~\ref{table:SP0}, and the useful cycles of phase are in listed in Table \ref{table:usecyc12}.}
\end{figure}
Figure~\ref{pp_pm_mm} shows the PDFs of the recovered $\beta_{-3}$ parameter for a one ppE parameter template family, with injections given by convergent, critical and asymptotic versions of cases (ii) and (iii). These PDFs are computed using a MCMC approach. The top panel of this figure shows the PDFs for $\beta_{-3}$ given an asymptotic injection, the middle panel given a critical injection, and the bottom panel given a convergent injection. The left and right panels correspond to injections with the same (left) or alternating (right) signs. When there is as much or more weight at $\beta_{-3} = 0$ in the PDF's as there was in the prior probability density, this indicates that GR is the preferred model. In our case, the prior probability for $\beta_{-3}$ is flat between $-5.0$ and $5.0$, and so the prior probability density at all points, including $\beta_{-3} = 0$, is $0.1$. When the posterior density at $\beta_{-3} =0$ is less than $0.1$, an alternative model is preferred.

Figure~\ref{pp_pm_mm} reveals several interesting facts. First, observe that in the convergent and in the asymptotic injection cases, the sign of the $\beta$s is irrelevant: in both cases most of the weight is outside $\beta_{-3} = 0$. Second, observe that in the convergent case, the second ppE term ($b=-2$) is very sub-dominant to the first term, and so its sign has little impact on the results. Third, observe that in the critical injection case, when the $\beta$s have alternating signs, the modified gravity effects partially cancel out, yielding a $\beta_{-3}$ PDF with non-negligible weight at the GR value.  It is clear from these studies that neglecting higher-order phase corrections can seriously bias our assessment of our ability to test GR with GW signals. For the ``Critical'' case, our ability to detect departures from GR is enhanced if the terms have the same sign, and diminished if the signs alternate.

%-----------------------------------------------------------------------------------
\section{Noise modeling and signal strength}
\label{sec:noise-SNR}

Most of our studies have been conducted on signals that do not have a noise realization explicitly added to the signal injection, although all analyses incorporate the noise spectrum of the detectors in the likelihood calculation. We chose not to include an explicit noise realization in order to expedite the calculation of the likelihood~\cite{Cornish:2010kf}, which then allows us to produce long Markov chains that fully explore the high dimensional parameter spaces. Unfortunately, our use of this technique has led some to question the reliability of our
results~\cite{Li:2011cg,Vallisneri:2011ts}. Here we show that those concerns are unfounded.

\begin{figure}[ht]
\begin{center}
\begin{tabular}{cc}
\epsfig{file=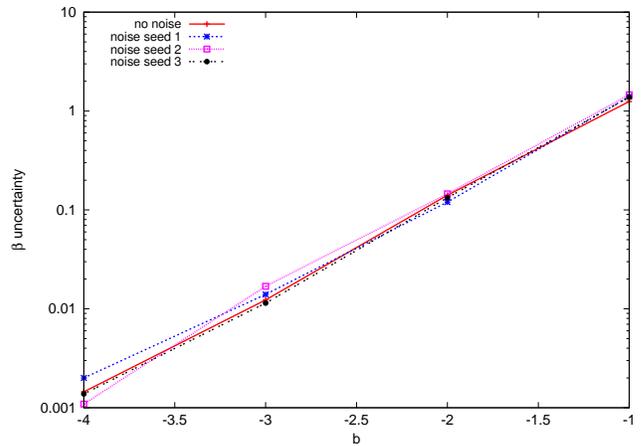, width=6cm,angle=-90} 
\end{tabular}
\end{center}
\vspace*{-0.2in}
\caption{\label{noise_BF} (Color Online) $(3\sigma)$-bounds on $\beta$ that can be inferred for different values of $b$, calculated from the PDF's of $\beta$ generated by recovering a GR signal with a ppE template. This plot shows the bounds for both a signal with no noise, and three that include Gaussian noise, generated from three different random seeds. The results are essentially identical. The signal parameters for this injection are in Table~\ref{table:SP0}.}
\end{figure}
The inclusion of noise in our signals has little effect on the conclusions we drew in our previous paper, as can be seen in  Fig.~\ref{noise_BF}. In this figure, we plot the $(3\sigma)$-bounds that we could place on the ppE phase parameters, if one has detected a NS-NS inspiral with SNR 15 that has no GR deviation. To calculate these bounds, we inject a GR signal and try to recover it using a single parameter ppE template, ie.~Eq.~\eqref{eq:phase-cor} with a single $\beta$. For any given value of $b$, we integrate out over all other parameters and take the standard deviation of the $\beta$ PDF as a $1\sigma$ bound. In other words, the curves show the upper limit of the magnitude a ppE parameter could be found to have, and still have the signal be consistent with GR. This plot shows that the bounds placed on the ppE parameters from a signal that includes an explicit noise realization are consistent with those found when no noise is added to the signal. That is, including an explicit noise realization does not affect the conclusions derived from a cheap-bound calculation with noise accounted for only through the detectors' noise spectrum in the likelihood.

\begin{figure}[ht]
\begin{center}
\begin{tabular}{cc}
\epsfig{file=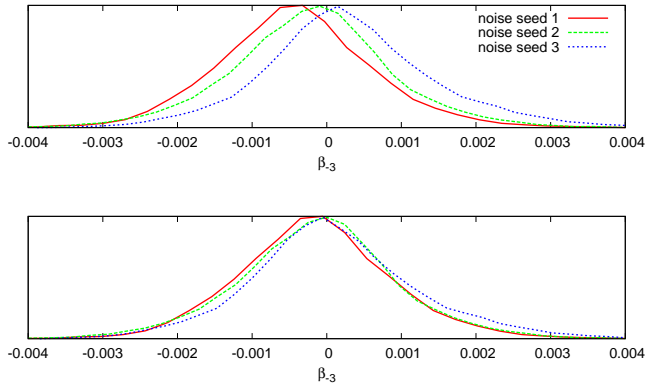, width=6cm,angle=-90} 
\end{tabular}
\end{center}
\vspace*{-0.2in}
\caption{\label{noise_3seeds} (Color Online) The top panel shows posterior distributions of $\beta$ recovered from three ppE injections, including noise in the injection. Each of the three signals was generated using a different random seed for the noise, but the same system parameters. The lower panel shows the same distributions, now with the best-fit  value of $\beta$ subtracted. This illustrates that, although noise affects the peak of the posterior distribution for a given parameter, it does not affect the uncertainty in that parameter. Thus the {\emph{cheap bounds}} of~\cite{Cornish:2011ys} are unaffected by the inclusion of noise.}
\end{figure}

To understand this result, it is useful to look at Figure~\ref{noise_3seeds}, which shows the recovered PDF's for the $\beta$ parameter from three different runs, each including noise generated with a different random seed. Since the injected signal was a GR NS-NS inspiral waveform with SNR 15, we would expect the $\beta$ PDF's to peak at zero. It is clear from this figure that, although the peak of the PDF is shifted by the inclusion of noise, the uncertainty in the recovery of this parameter, i.e. the spread of the distribution, is not affected. This concept has been explored before, in \cite{Nissanke:2009kt} and \cite{Vallisneri:2011ts}. In  \cite{Nissanke:2009kt}, the authors argue that when discussing our ability to measure system parameters in general, and not for a particular case, what we really want to do is examine the \emph{noise-averaged} uncertainties in these parameters. That is, we are interested in how well we can measure parameters when averaged over many specific realizations of the noise. The authors show that the noise-averaged uncertainties are the same as the uncertainties calculated with zero noise injected into the signal. In \cite{Vallisneri:2011ts} it is argued that the specific noise realization will affect our parameter estimation, and while this is technically true, we have shown in this section that the overall effect is minimal. In any case, for the type of analysis that we want to do in the rest of this paper, the reasoning of \cite{Nissanke:2009kt} applies, and so we do not inject an explicit noise realization for any of our analyses in the other sections. It has also been claimed in~\cite{Li:2011cg} that simulated data that only includes a signal injection, ie.~that does not include a noise realizations, will necessarily lead to posterior distributions for the system parameters that are Gaussian. This is patently false, as can easily be demonstrated by analytically calculating the posterior distribution for a signal of the form $(d_0/d) \cos(2 \pi f t)$, which leads to a highly non-Gaussian distribution in the distance $d$.

\begin{figure}[ht]
\begin{center}
\begin{tabular}{cc}
\epsfig{file=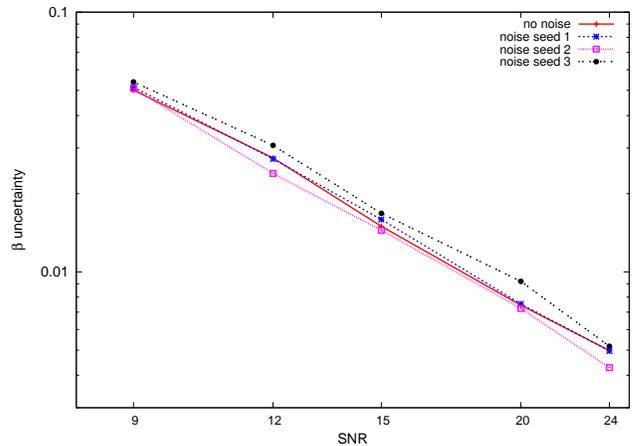, width=6cm,angle=-90} 
\end{tabular}
\end{center}
\vspace*{-0.2in}
\caption{\label{SNRsigma} (Color Online) $(3\sigma)$-bounds on $\beta$  for $b = -1.0$, calculated from the PDF's of $\beta$ generated by recovering a GR signal with a ppE template. This plot shows the linear relationship between the bounds on $\beta$ and the SNR of the signal. There are four lines shown - one for a signal that had no noise injected, and three for signals that had noise injected, each with a different random seed. The results are essentially identical. The signal parameters for this injection are in Table~\ref{table:SP0}.}
\end{figure}

Obviously, signals with high SNR will be better for testing GR, as they are better for any type of GW data analysis. When discussing how well GR can be tested using GW detections, the highest-SNR events are the ones that will lead to the strongest constraints. In our previous paper, we analyzed signals with SNR $\sim 20$, which would be considered a high SNR detection by the LIGO detectors. It is irrelevant, however, that most signals will probably have SNRs in the low $10$s. There will always be one signal with highest SNR, and this is likely to be above $15$. It is therefore still useful to study GR tests assuming detections with SNRs $\sim 20$, as it is not a hopeless proposition that we will have this type of event in our GW catalog. Throughout the rest of this paper, however, we have taken a more pessimistic tack, and restricted ourselves to analyzing signals with SNR $\sim 10 - 12$. The results follow the theoretical linear scaling with SNR~\cite{Cornish:2010kf} down to values of the SNR that are close to the detection threshold, which for this system was found to be ${\rm SNR} \sim 7.5$. This scaling is shown in Figure~\ref{SNRsigma}.

%
%%%%%%%%%%%%%%%%%%%%%%%%%%%%%%%%%%
\section{Optimal Model Selection}
\label{sec:optimal-fitting}
\begin{figure*}[ht]
\begin{center}
\begin{tabular}{cc}
\epsfig{file=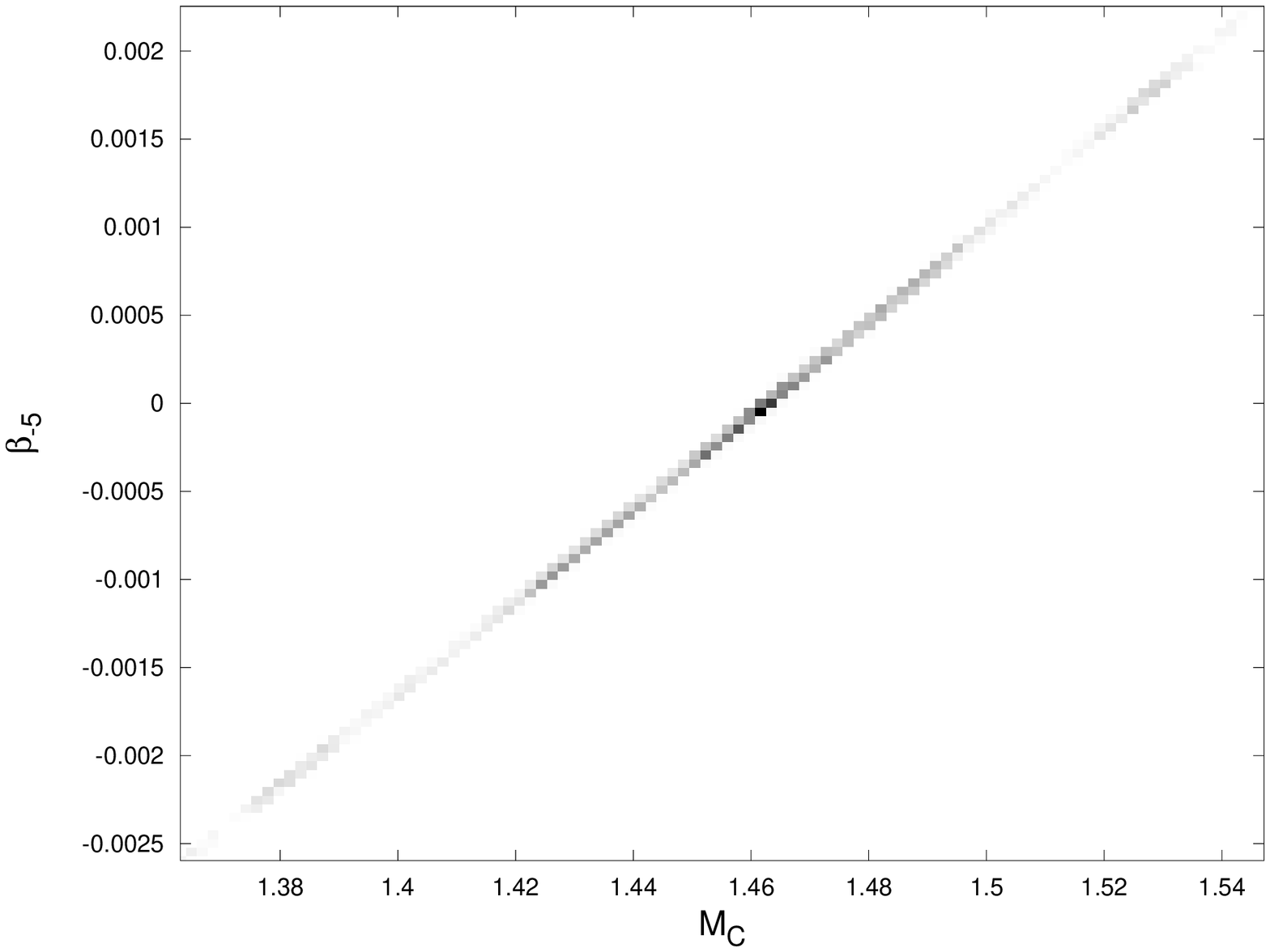, width=9.0cm,angle=0}  &
\epsfig{file=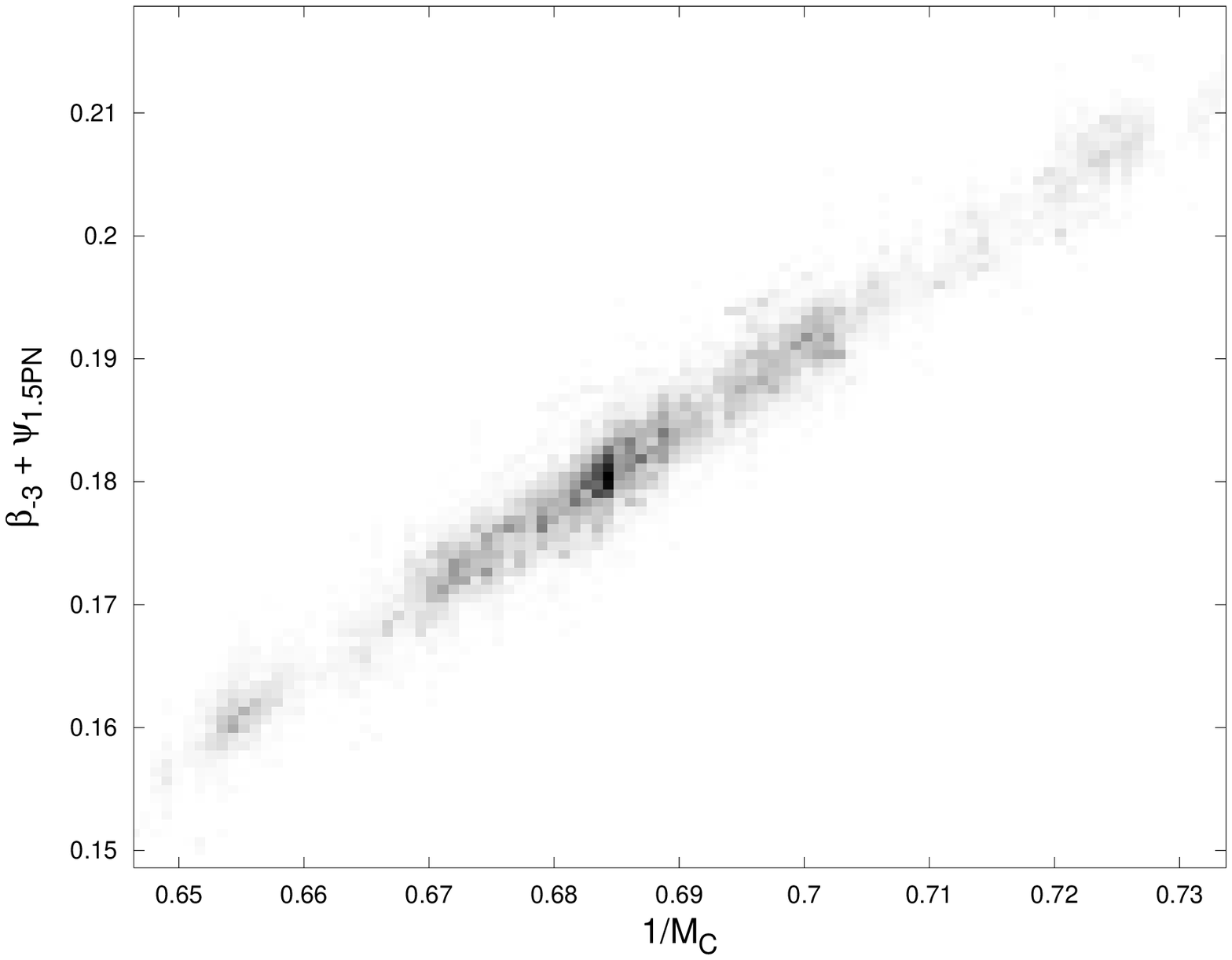, width=9.0cm,angle=0} 
\vspace*{-0.2in}
\end{tabular}
\end{center}
\caption{\label{beta1Mcorr} Correlation between the $\beta_{-5}$ ppE parameter and the chirp mass (left panel) and the $\beta_{-3}$ parameter and the inverse chirp mass (right panel) for an injection including two PN phase terms as well as two ppE phase corrections. The parameters are restricted only by their prior ranges. }
\end{figure*}

We have seen that it is important to consider multi-term ppE signal injections when assessing the bounds we will be able to place on alternative gravity theories. The question still remains, however, as to what type of templates we should use to recover such signals. In this section we address this question by showing first that adding too many parameters to the templates is counter-productive. Then we determine the optimal ppE template family to detect departures from GR described by the more realistic multi-term ppE signal injection model.

%-----------------------------------------------------------------------------------
\subsection{Overfitting}
\label{overfitting}

One may consider using a ppE template with many ppE phase and amplitude terms in the sums of Eq.~\eqref{eq:gen-ppE}. For example, one could include as many ppE phase terms as there are in the GR PN series, but this is far from ideal. The reason is clear: if we include the same number of free ppE parameters in our phase model as we have phase terms that are functions of system parameters, then there is no way to constrain any of them. In other words, the ppE phase terms will have a $100\%$ correlation with the standard GR system parameters that form the coefficients of the GR PN phase. 

As a simple example, consider the possibility of detecting a non-GR signal that includes ppE corrections at $b=-5$ (a so-called {\emph{Newtonian}} ppE correction) and $b=-3$ (a 1PN ppE correction). We will truncate our injection at 1PN order for this example, which implies that the GW phase contains two standard PN terms that are functions of the system parameters, and two free ppE terms. Figure~\ref{beta1Mcorr} shows that there is a $100\%$ correlation between these PN and ppE parameters. 

These types of correlations are commonly encountered in GW data analysis, but they may not be widely appreciated by theoretical model builders.
We can understand this correlation analytically as follows. Let us write the simplified ppE template Fourier phase $\Psi_{\ppe}(f)$ as follows
\begin{align}
\Psi_{\ppe}(f) &= \left[\frac{3}{128} \left(\pi {\cal{M}}\right)^{-5/3} + \beta_{-5} \left(\pi {\cal{M}}\right)^{-5/3} \right]f^{-5/3} 
\nonumber \\
&+ \left[ \frac{3}{128 \eta^{2/5} \pi {\cal{M}}} \left(\frac{3715}{756} + \frac{55}{9} \eta \right) + \frac{\beta_{-3}}{\pi {\cal{M}}} \right]  f^{-1}\,.
\end{align}
where we have expanded out the definition of $u$. Clearly, we can rescale $\beta_{-5}$ by a constant and $\beta_{-3}$ by a function of $\eta$ to recover the same value of the Fourier phase, thus showing a direct correlation between these parameters. Figure~\ref{beta1Mcorr} demonstrates how such a correlation manifests itself in the posterior distributions.

This argument can be extended to whatever PN order we choose. If we include the same number of ppE terms as PN terms in our model, then we will not be able to place bounds on {\emph{any}} parameter, let alone use the results as a test of GR. It is also true, however, that ppE models that include more ppE terms will be able to achieve a better overall fit of whatever signal we happen to detect, just as any model with extra parameters can typically fit data better than a simpler model. In the next section we explore the tradeoff between these two effects. We also attempt to determine what types of signals are best to analyze using more complex ppE models, and what types are better served with a simple ppE model.  

%-----------------------------------------------------------------------------------
\subsection{Parsimonious Fitting: Detecting and Characterizing non-GR signals}

\begin{table*}[ht]
\centering  % used for centering table
\begin{tabular}{l c c c c c c c c c c c c} % centered columns (4 columns)
\hline\hline                        %inserts double horizontal lines
Source &$\alpha$ & $\phi_L$ & $\phi_c$ & $m_1(M_{\bigodot})$ & $m_2(M_{\bigodot})$ & $\log(D_L) ({\rm{Mpc}})$ & $t_c$ & $\delta$ &$\theta_L$ & $\beta_{-3}$ & $\beta_{-2}$ & $\beta_{-1}$ \\ [0.5ex] % inserts table 
%heading
\hline                  % inserts single horizontal line
Convergent & 1.42   & 2.5 & 0.8 &1.42 &1.73 & 3.83 &3.5 &0.87 &0.43 &0.003   &0.003  &0.003 \\ 
Critical          & 1.42   & 2.5 & 0.8 &1.52 &1.33 & 3.9 &3.5 &0.87 &0.43 &0.0006 &0.018 &0.54  \\
Asymptotic   & 1.42   & 2.5 & 0.8 &2.04 &1.34 & 3.86 &3.5 &0.87 &0.43 &0.0007  &0.07  &7.0  \\ [1ex]      % [1ex] adds vertical space
\hline %inserts single line
\end{tabular}
\caption{\label{table:SP57} Source parameters for Figures \ref{BFsub} and \ref{bpdf}. The $\beta_b$ values listed are for a particular case - the ratio between different $\beta_b$ values was kept constant for each injected signal. The ratio for convergent was $\times 1.0$, critical was $\times 30$, and asymptotic was $\times 100$.}
% is used to refer this table in the text
\end{table*}

Let us now study what type of ppE templates are best suited for detecting a GR deviation. In particular, let us examine whether using one-term or two-term ppE templates works better.  For this analysis, we inject ppE signals containing three phase terms, and attempt to recover them using one- and two-parameter ppE templates. We calculate Bayes factors between the ppE models against the GR model to see which model is best suited to detecting departures from GR. Because of our strong prior belief in the validity of GR, a Bayes factor significantly greater than unity would be necessary to convince us that a new theory of gravity is needed.

\begin{figure*}[htb]
\begin{center}
\epsfig{file=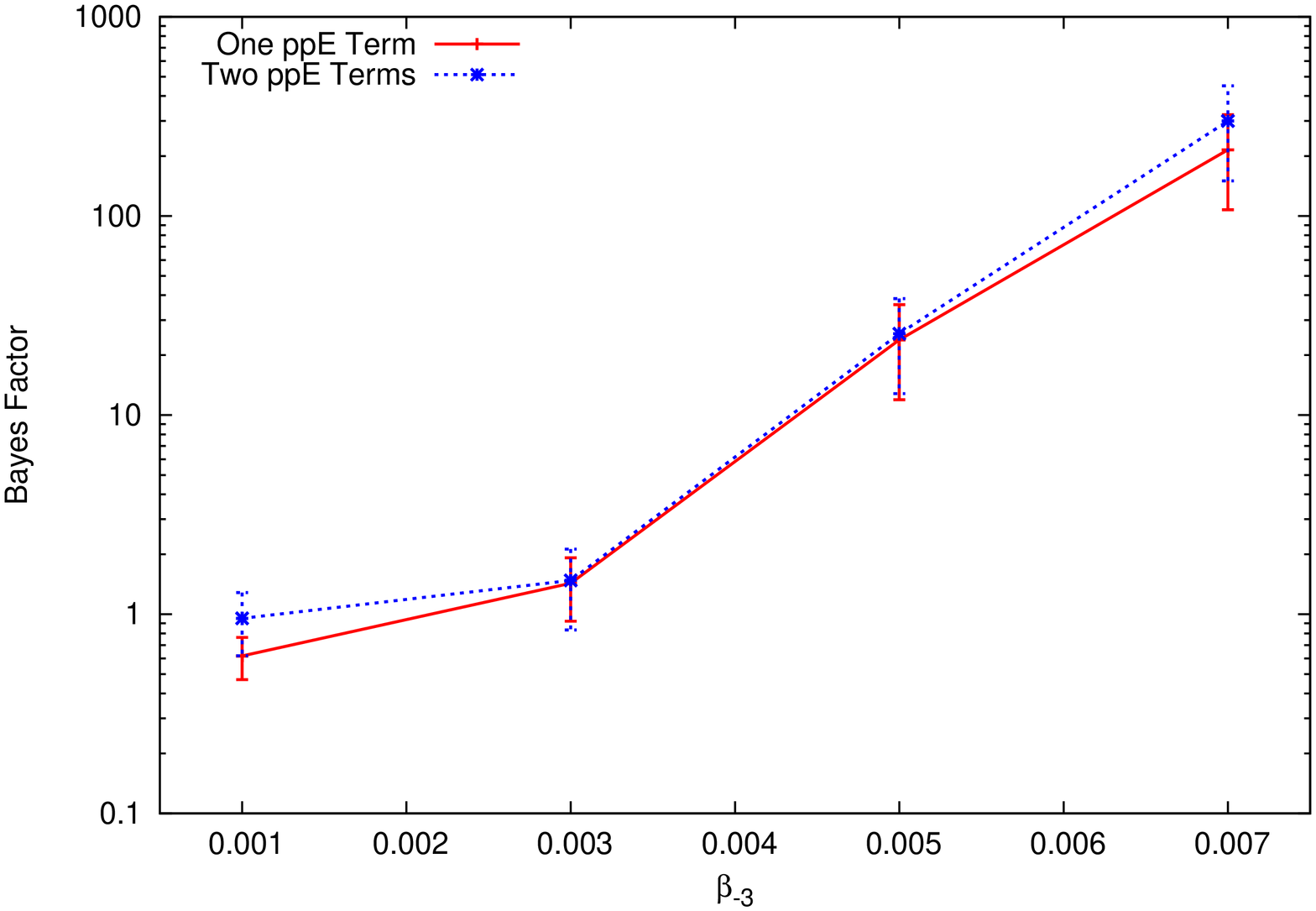, width=6cm,angle=-90} \\
\epsfig{file=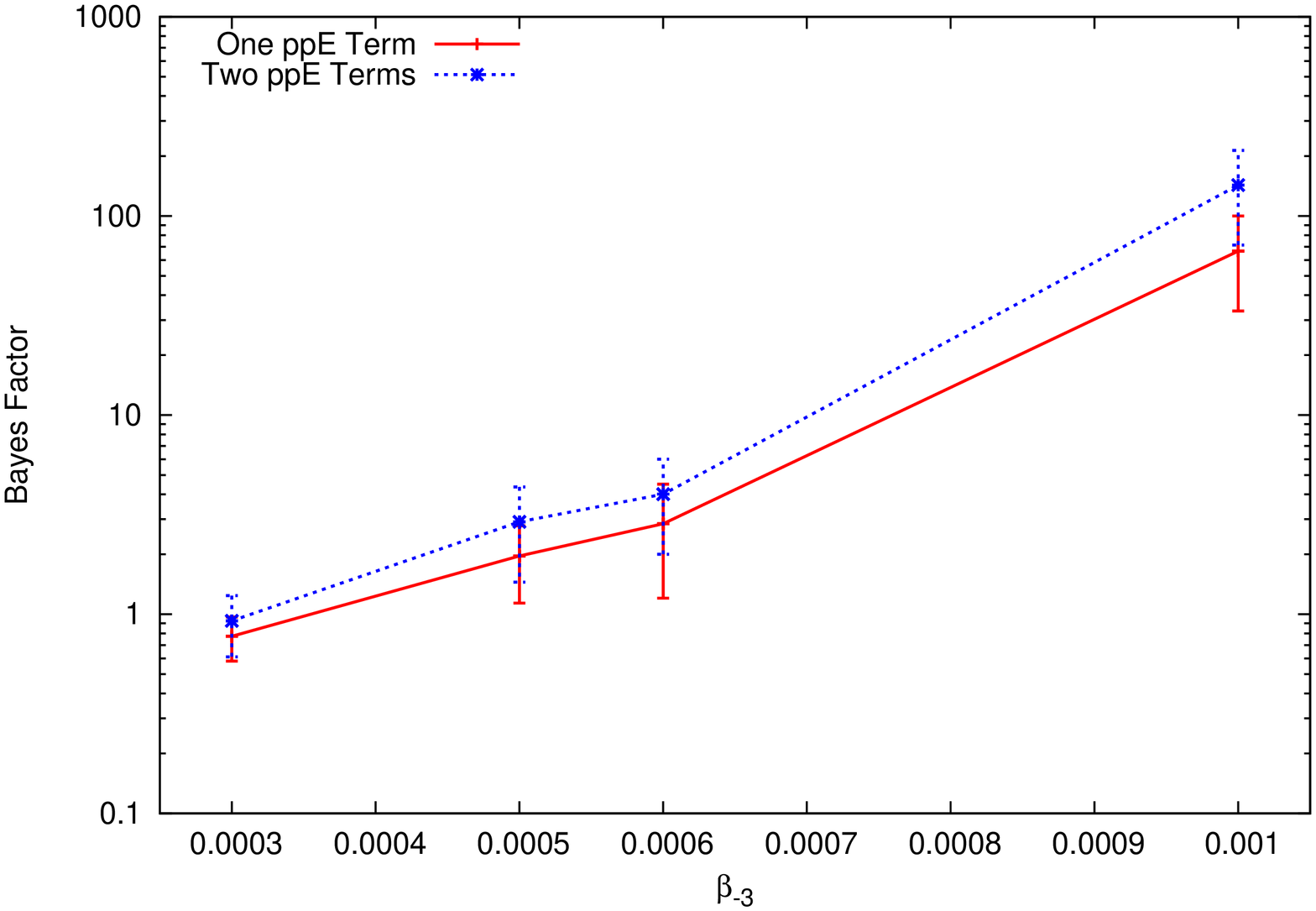, width=6cm,angle=-90} \\
\epsfig{file=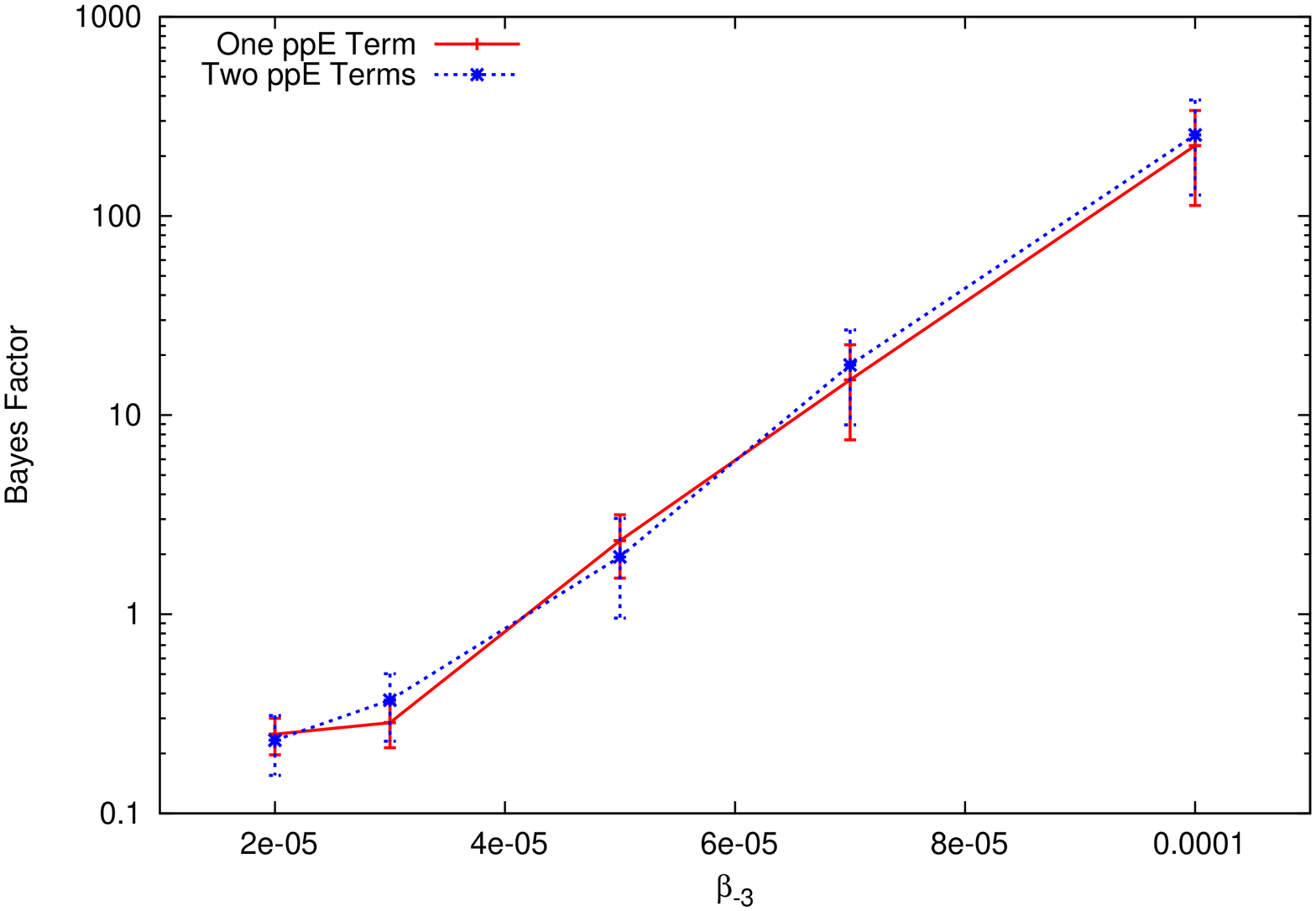, width=6cm,angle=-90}
\end{center}
\vspace*{-0.2in}
\caption{\label{BFsub} (Color Online) Bayes factors for one-term (solid red) and two-term (dashed blue) ppE templates for a convergent (top-left), critical (top-right) and asymptotic (bottom) ppE injection as a function of the injected value of $\beta_{-3}$. System parameters are listed in Table~\ref{table:SP57}, and useful cycles of phase in Table \ref{table:usecyc57}. In the convergent and asymptotic cases, both models perform equally well at detecting a deviation from GR. In the critical case, the two-term model slightly out-performs the one-term model. } 
\end{figure*}

Let us then consider three different ppE injections, starting at 1 PN order ($b=-3,-2,-1$), a convergent, a critical and an asymptotic one, each for a NS-NS inspiral, with parameters listed in Table~\ref{table:SP57}  . We explore these simulated signals with a MCMC algorithm, using a one- and a two-term ppE model. The one-term ppE models are allowed to choose between phase exponents $b=-3$ and $b=-2$,  while the two-term models are allowed to choose between the pairs $(-3,-2)$ and $(-2,-1)$ - {\it i.e.} the two terms must differ by a single power of $u$, and models with exponents $(-3, -1)$ are not allowed.

The Bayes factors between the one-term ppE model and GR (red solid curve) and between the two-term ppE model and GR (blue dashed curve) are shown in Fig.~\ref{BFsub} as a function of the injected value of $\beta_{-3}$ for a convergent (top-left panel), critical (top-right panel) and asymptotic (bottom panel) injection. These Bayes Factors are again calculated using the Savage-Dicke density ratio. Calculating the posterior density at a $\beta_i = 0$ from a Markov chain involves counting the number of points in the chain that fall within the histogram bin containing $\beta_i = 0$, and so the error bars reflect the counting error involved in this process, as well as the spread in BF values calculated from multiple MCMC runs on the same signal but with different random seeds. Observe that the only injections for which two-term ppE templates consistently outperform one-term ppE templates are the critical ones. Even in this case, however, the preference is not large; the curves track each other very well in all cases. Therefore, our results indicate that the one-term ppE templates are sufficient for searching for deviation from GR in GW data.

\begin{figure}[ht]
\begin{center}
\begin{tabular}{cc}
\epsfig{file=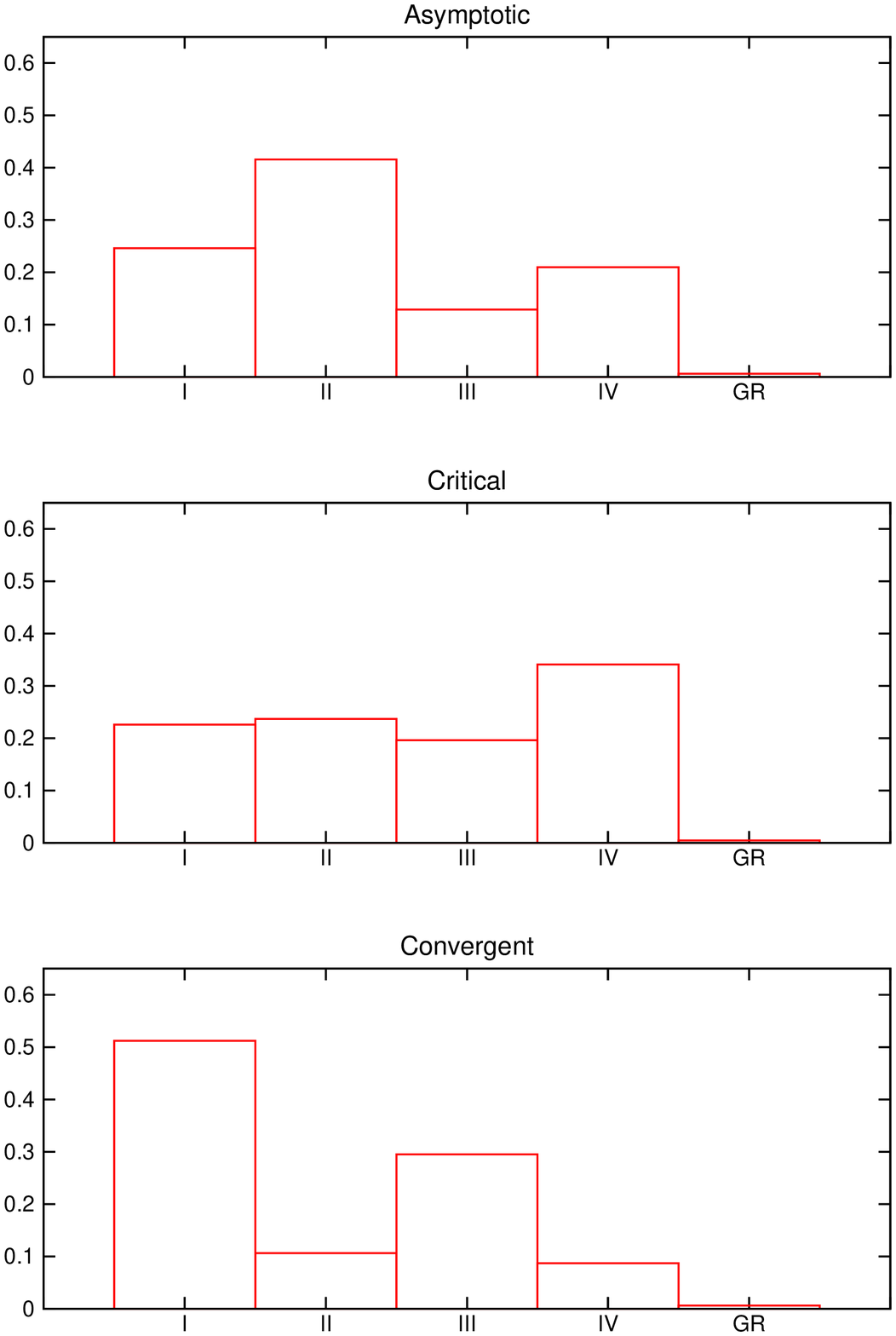, width=6cm,angle=0} 
\end{tabular}
\end{center}
\caption{\label{bpdf}  Posterior distributions for the four different ppE models, generated by RJMCMC. The top two panels show the distribution for a convergent injection, the middle two for a critical injection, and the bottom two for an asymptotic injection. All systems are NS-NS binaries with Bayes Factors of 100 favoring ppE over GR. System parameters are in Table \ref{table:SP57}. Model I has $b=-3$, model II has $b=-2$, model III has $b=-3$ \emph{and} $b=-2$, and model IV has $b=-2$ \emph{and} $b=-1$. The y axis shows the percentage of iterations that the chain spent in each model, and the Bayes Factors between two models are simply the ratios of the percentages. Because the Bayes Factors are not large enough, these results indicate that we would not be able to make confident statements about the type of non-GR signal we had observed with this type of analysis.}
\end{figure}

Once a deviation from GR has been definitively detected, the next step is to learn as much about the signal as possible, in order to give theorists as much guidance as possible in their attempts to build an alternative theory of gravity. The information we could hope to extract from the type of analysis we have described in this paper is the structure of the series of phase corrections - do they enter at a certain PN power and then fade away? Or do they enter at that power and grow more important at higher orders in the expansion? Figure~\ref{bpdf} plots the posterior distribution of the five models under consideration derived using a RJMCMC~\cite{cornish:083006} analysis. In RJMCMC, moves are proposed between models of different dimensionality according to the Metropolis-Hastings ratio:
\be
\alpha = {\rm min} \left \{    1, \frac{p(\vec{\lambda)}_Y p(s|\vec{\lambda}_Y) q(\vec{u}_Y)} {p(\vec{\lambda)}_X p(s|\vec{\lambda}_X) q(\vec{u}_X)} |\mathbb{J}|  \right \}
\ee
Here, model X and model Y differ by some number of parameters, $q(\vec{u})$ is the distribution for random numbers chosen to generate the extra parameters, and $|\mathbb{J}|$ is the Jacobian of the two sets of parameters, which compensates for the difference in dimensionality. When using this Hastings ratio as an acceptance probability, we can allow our chains to explore the full space of allowed ppE models, both one- and two-term families, and use these to generate PDF's for the models themselves. The ratio of the heights of the PDF for model X and model Y is equal to the Bayes Factor between X and Y.

\begin{table}[ht]
%\centering  % used for centering table
\begin{tabular}{l c c c} % centered columns (4 columns)
\hline\hline                        %inserts double horizontal lines
Signal & $\phi_{-3}$ & $\phi_{-2}$& $\phi_{-1}$ \\ [0.5ex] % inserts table 
%heading
\hline                  % inserts single horizontal line
Convergant & 0.109 & 0.008 & 0.0005  \\ % inserting body of the table
Critical          & 0.024 & 0.051 & 0.085  \\
Asymptotic   & 0.024 & 0.181 & 1.047 \\ [1ex]      % [1ex] adds vertical space
\hline %inserts single line
\end{tabular}
 % is used to refer this table in the text
\caption{\label{table:usecyc57} Number of useful cycles from the different injected ppE terms - Fig. \ref{BFsub}}
\end{table}

To generate Figure ~\ref{bpdf}, we have run a RJMCMC search on three different types of signals - one convergent, one critical, and one asymptotic - and plotted the number of iterations that the chains spent in each of the five different models. These five models include two ppE models with only one phase correction, ($b=-3$ and $b=-2$), two ppE models with two phase corrections, ($b=-3 + b=-2$ and $b=-2 + b=-1$), and GR. We find that, although there are some slight differences between the different models, in all cases we cannot draw meaningful distinctions between the different ppE models. The strongest Bayes Factor between two models is in the convergent case, where the Bayes Factor between the $b=-3$ only model and the $b=-2$ only model is $\approx 5$. While this does show some preference for the first model, it is not a strong preference, and so we would not want to use this result to draw conclusions about the underlying theory of gravity. In summary - even though these signals are clearly differentiable from GR (all have Bayes Factors of $\approx 100$), the four different ppE models perform almost as well in fitting the signal. This means that if we hope to gain more information about the underlying nature of an alternative gravity theory, we would need higher SNR signals and/or multiple detections. On a more hopeful note, it means that our ability to detect a deviation from GR is not strongly dependent on which particular ppE template we choose to use in our analysis.

%%%%%%%%%%%%%%%%%%%%%%%%%%%%%%%%%%
\section{Future Directions}
\label{sec:conclusions}

In this paper, we have investigated the effects of using more realistic non-GR injections to investigate our ability to test GR using GW signals. We have found that the inclusion of noise in our analysis does not significantly affect our results, but that the failure to include higher-order deviations from GR in the phase of the injected signal can bias them. We have also determined that one-parameter ppE template families are best for detecting deviations from GR, at least for the simple cases investigated here.

The main direction of future work will be in determining how analyzing more \emph{astrophysically} realistic systems affects our ability to test GR. That is, systems that incorporate not only more complicated deviations from GR, such as we examined in this paper, but that also include some of the messiness we know will exist in real systems in our universe. For instance, if we were analyzing systems that merge within the aLIGO frequency band, we would need to include the merger and ringdown parts of the waveforms in our injections. If we then performed a Bayesian model selection between ppE and GR \emph{inspiral-only} templates, it is entirely possible that the ppE templates would win over the GR ones, simply by being able to fit more of the power in the non-inspiral signal. It is also possible that the presence of accretion disks in real BH systems could alter the GW signature of the systems enough that we could mistakenly claim to have detected a deviation from GR. We will examine these potential sources of systematic error in a future paper.

%%%%%%%%%%%%%%%%%%%%%%%%%%%%%%%%%%
%%%%%%%%%%%%%%%%%%%%%%%%%%%%%%%%%%
%%%%%%%%%%%%%%%%%%%%%%%%%%%%%%%%%%
\acknowledgments

We thank Michelle Vallisneri for his useful comments and suggestions.
N.~C. and L.~S.~acknowledge support from the NSF Award PHY-1205993  and NASA grant
NNX10AH15G. N.~Y.~acknowledges support from NSF grant PHY-1114374 and 
NASA grant NNX11AI49G, under sub-award 00001944.

%%%%%%%%%%%%%%%%%%%%%%%%%%%%%%%%%%

%%%%%%%%%%%%%%%%%%%%%%%%%%%%%%%%%%
%%%%%%%%%%%%%%%%%%%%%%%%%%%%%%%%%%
%%%%%%%%%%%%%%%%%%%%%%%%%%%%%%%%%%
\bibliography{master.bib}

\begin{thebibliography}{35}
\expandafter\ifx\csname natexlab\endcsname\relax\def\natexlab#1{#1}\fi
\expandafter\ifx\csname bibnamefont\endcsname\relax
  \def\bibnamefont#1{#1}\fi
\expandafter\ifx\csname bibfnamefont\endcsname\relax
  \def\bibfnamefont#1{#1}\fi
\expandafter\ifx\csname citenamefont\endcsname\relax
  \def\citenamefont#1{#1}\fi
\expandafter\ifx\csname url\endcsname\relax
  \def\url#1{\texttt{#1}}\fi
\expandafter\ifx\csname urlprefix\endcsname\relax\def\urlprefix{URL }\fi
\providecommand{\bibinfo}[2]{#2}
\providecommand{\eprint}[2][]{\url{#2}}

\bibitem[{\citenamefont{Will}(2006)}]{lrr-2006-3}
\bibinfo{author}{\bibfnamefont{C.~M.} \bibnamefont{Will}},
  \bibinfo{journal}{Living Reviews in Relativity} \textbf{\bibinfo{volume}{9}}
  (\bibinfo{year}{2006}),
  \urlprefix\url{http://www.livingreviews.org/lrr-2006-3}.

\bibitem[{\citenamefont{Kramer et~al.}(2006)}]{Kramer:2006nb}
\bibinfo{author}{\bibfnamefont{M.}~\bibnamefont{Kramer}} \bibnamefont{et~al.},
  \bibinfo{journal}{Science} \textbf{\bibinfo{volume}{314}},
  \bibinfo{pages}{97} (\bibinfo{year}{2006}), \eprint{astro-ph/0609417}.

\bibitem[{\citenamefont{Alexander and Yunes}(2009)}]{Alexander:2009tp}
\bibinfo{author}{\bibfnamefont{S.}~\bibnamefont{Alexander}} \bibnamefont{and}
  \bibinfo{author}{\bibfnamefont{N.}~\bibnamefont{Yunes}},
  \bibinfo{journal}{Phys. Rept.} \textbf{\bibinfo{volume}{480}},
  \bibinfo{pages}{1} (\bibinfo{year}{2009}), \eprint{0907.2562}.

\bibitem[{\citenamefont{Will}(1994)}]{Will:1994fb}
\bibinfo{author}{\bibfnamefont{C.~M.} \bibnamefont{Will}},
  \bibinfo{journal}{Phys. Rev.} \textbf{\bibinfo{volume}{D50}},
  \bibinfo{pages}{6058} (\bibinfo{year}{1994}), \eprint{gr-qc/9406022}.

\bibitem[{\citenamefont{Scharre and Will}(2002)}]{Scharre:2001hn}
\bibinfo{author}{\bibfnamefont{P.~D.} \bibnamefont{Scharre}} \bibnamefont{and}
  \bibinfo{author}{\bibfnamefont{C.~M.} \bibnamefont{Will}},
  \bibinfo{journal}{Phys. Rev.} \textbf{\bibinfo{volume}{D65}},
  \bibinfo{pages}{042002} (\bibinfo{year}{2002}), \eprint{gr-qc/0109044}.

\bibitem[{\citenamefont{Will and Yunes}(2004)}]{Will:2004xi}
\bibinfo{author}{\bibfnamefont{C.~M.} \bibnamefont{Will}} \bibnamefont{and}
  \bibinfo{author}{\bibfnamefont{N.}~\bibnamefont{Yunes}},
  \bibinfo{journal}{Class. Quant. Grav.} \textbf{\bibinfo{volume}{21}},
  \bibinfo{pages}{4367} (\bibinfo{year}{2004}), \eprint{gr-qc/0403100}.

\bibitem[{\citenamefont{Berti et~al.}(2005)\citenamefont{Berti, Buonanno, and
  Will}}]{Berti:2005qd}
\bibinfo{author}{\bibfnamefont{E.}~\bibnamefont{Berti}},
  \bibinfo{author}{\bibfnamefont{A.}~\bibnamefont{Buonanno}}, \bibnamefont{and}
  \bibinfo{author}{\bibfnamefont{C.~M.} \bibnamefont{Will}},
  \bibinfo{journal}{Class. Quant. Grav.} \textbf{\bibinfo{volume}{22}},
  \bibinfo{pages}{S943} (\bibinfo{year}{2005}), \eprint{gr-qc/0504017}.

\bibitem[{\citenamefont{Yagi and Tanaka}(2009)}]{Yagi:2009zm}
\bibinfo{author}{\bibfnamefont{K.}~\bibnamefont{Yagi}} \bibnamefont{and}
  \bibinfo{author}{\bibfnamefont{T.}~\bibnamefont{Tanaka}}
  (\bibinfo{year}{2009}), \eprint{0906.4269}.

\bibitem[{\citenamefont{Will}(1998)}]{Will:1997bb}
\bibinfo{author}{\bibfnamefont{C.~M.} \bibnamefont{Will}},
  \bibinfo{journal}{Phys. Rev.} \textbf{\bibinfo{volume}{D57}},
  \bibinfo{pages}{2061} (\bibinfo{year}{1998}), \eprint{gr-qc/9709011}.

\bibitem[{\citenamefont{Stavridis and Will}(2009)}]{Stavridis:2009mb}
\bibinfo{author}{\bibfnamefont{A.}~\bibnamefont{Stavridis}} \bibnamefont{and}
  \bibinfo{author}{\bibfnamefont{C.~M.} \bibnamefont{Will}},
  \bibinfo{journal}{Phys. Rev.} \textbf{\bibinfo{volume}{D80}},
  \bibinfo{pages}{044002} (\bibinfo{year}{2009}), \eprint{0906.3602}.

\bibitem[{\citenamefont{Arun and Will}(2009)}]{Arun:2009pq}
\bibinfo{author}{\bibfnamefont{K.~G.} \bibnamefont{Arun}} \bibnamefont{and}
  \bibinfo{author}{\bibfnamefont{C.~M.} \bibnamefont{Will}},
  \bibinfo{journal}{Class. Quant. Grav.} \textbf{\bibinfo{volume}{26}},
  \bibinfo{pages}{155002} (\bibinfo{year}{2009}), \eprint{0904.1190}.

\bibitem[{\citenamefont{Keppel and Ajith}(2010)}]{Keppel:2010qu}
\bibinfo{author}{\bibfnamefont{D.}~\bibnamefont{Keppel}} \bibnamefont{and}
  \bibinfo{author}{\bibfnamefont{P.}~\bibnamefont{Ajith}},
  \bibinfo{journal}{Phys. Rev.} \textbf{\bibinfo{volume}{D82}},
  \bibinfo{pages}{122001} (\bibinfo{year}{2010}), \eprint{1004.0284}.

\bibitem[{\citenamefont{Alexander et~al.}(2008)\citenamefont{Alexander, Finn,
  and Yunes}}]{Alexander:2007:gwp}
\bibinfo{author}{\bibfnamefont{S.}~\bibnamefont{Alexander}},
  \bibinfo{author}{\bibfnamefont{L.~S.} \bibnamefont{Finn}}, \bibnamefont{and}
  \bibinfo{author}{\bibfnamefont{N.}~\bibnamefont{Yunes}},
  \bibinfo{journal}{Phys. Rev. D} \textbf{\bibinfo{volume}{78}},
  \bibinfo{pages}{066005} (\bibinfo{year}{2008}), \eprint{0712.2542}.

\bibitem[{\citenamefont{Yunes et~al.}(2010)\citenamefont{Yunes, O'Shaughnessy,
  Owen, and Alexander}}]{Yunes:2010yf}
\bibinfo{author}{\bibfnamefont{N.}~\bibnamefont{Yunes}},
  \bibinfo{author}{\bibfnamefont{R.}~\bibnamefont{O'Shaughnessy}},
  \bibinfo{author}{\bibfnamefont{B.~J.} \bibnamefont{Owen}}, \bibnamefont{and}
  \bibinfo{author}{\bibfnamefont{S.}~\bibnamefont{Alexander}},
  \bibinfo{journal}{Phys. Rev.} \textbf{\bibinfo{volume}{D82}},
  \bibinfo{pages}{064017} (\bibinfo{year}{2010}), \eprint{1005.3310}.

\bibitem[{\citenamefont{Yunes and
  Pretorius}(2009{\natexlab{a}})}]{Yunes:2009hc}
\bibinfo{author}{\bibfnamefont{N.}~\bibnamefont{Yunes}} \bibnamefont{and}
  \bibinfo{author}{\bibfnamefont{F.}~\bibnamefont{Pretorius}},
  \bibinfo{journal}{Physical Review D (Particles, Fields, Gravitation, and
  Cosmology)} \textbf{\bibinfo{volume}{79}}, \bibinfo{eid}{084043}
  (pages~\bibinfo{numpages}{14}) (\bibinfo{year}{2009}{\natexlab{a}}),
  \urlprefix\url{http://link.aps.org/abstract/PRD/v79/e084043}.

\bibitem[{\citenamefont{Sopuerta and Yunes}(2009)}]{Sopuerta:2009iy}
\bibinfo{author}{\bibfnamefont{C.~F.} \bibnamefont{Sopuerta}} \bibnamefont{and}
  \bibinfo{author}{\bibfnamefont{N.}~\bibnamefont{Yunes}},
  \bibinfo{journal}{Physical Review D (Particles, Fields, Gravitation, and
  Cosmology)} \textbf{\bibinfo{volume}{80}}, \bibinfo{eid}{064006}
  (pages~\bibinfo{numpages}{24}) (\bibinfo{year}{2009}),
  \urlprefix\url{http://link.aps.org/abstract/PRD/v80/e064006}.

\bibitem[{\citenamefont{Yagi et~al.}(2012)\citenamefont{Yagi, Yunes, and
  Tanaka}}]{Yagi:2012vf}
\bibinfo{author}{\bibfnamefont{K.}~\bibnamefont{Yagi}},
  \bibinfo{author}{\bibfnamefont{N.}~\bibnamefont{Yunes}}, \bibnamefont{and}
  \bibinfo{author}{\bibfnamefont{T.}~\bibnamefont{Tanaka}}
  (\bibinfo{year}{2012}), \eprint{1208.5102}.

\bibitem[{\citenamefont{Yunes et~al.}(2009)\citenamefont{Yunes, Pretorius, and
  Spergel}}]{Yunes:2009bv}
\bibinfo{author}{\bibfnamefont{N.}~\bibnamefont{Yunes}},
  \bibinfo{author}{\bibfnamefont{F.}~\bibnamefont{Pretorius}},
  \bibnamefont{and} \bibinfo{author}{\bibfnamefont{D.}~\bibnamefont{Spergel}}
  (\bibinfo{year}{2009}), \eprint{0912.2724}.

\bibitem[{\citenamefont{Bekenstein}(2004)}]{Bekenstein:2004ne}
\bibinfo{author}{\bibfnamefont{J.~D.} \bibnamefont{Bekenstein}},
  \bibinfo{journal}{Phys. Rev.} \textbf{\bibinfo{volume}{D70}},
  \bibinfo{pages}{083509} (\bibinfo{year}{2004}), \eprint{astro-ph/0403694}.

\bibitem[{\citenamefont{Arun et~al.}(2006{\natexlab{a}})\citenamefont{Arun,
  Iyer, Qusailah, and Sathyaprakash}}]{Arun:2006hn}
\bibinfo{author}{\bibfnamefont{K.~G.} \bibnamefont{Arun}},
  \bibinfo{author}{\bibfnamefont{B.~R.} \bibnamefont{Iyer}},
  \bibinfo{author}{\bibfnamefont{M.~S.~S.} \bibnamefont{Qusailah}},
  \bibnamefont{and} \bibinfo{author}{\bibfnamefont{B.~S.}
  \bibnamefont{Sathyaprakash}}, \bibinfo{journal}{Phys. Rev.}
  \textbf{\bibinfo{volume}{D74}}, \bibinfo{pages}{024006}
  (\bibinfo{year}{2006}{\natexlab{a}}), \eprint{gr-qc/0604067}.

\bibitem[{\citenamefont{Arun et~al.}(2006{\natexlab{b}})\citenamefont{Arun,
  Iyer, Qusailah, and Sathyaprakash}}]{Arun:2006yw}
\bibinfo{author}{\bibfnamefont{K.~G.} \bibnamefont{Arun}},
  \bibinfo{author}{\bibfnamefont{B.~R.} \bibnamefont{Iyer}},
  \bibinfo{author}{\bibfnamefont{M.~S.~S.} \bibnamefont{Qusailah}},
  \bibnamefont{and} \bibinfo{author}{\bibfnamefont{B.~S.}
  \bibnamefont{Sathyaprakash}}, \bibinfo{journal}{Class. Quant. Grav.}
  \textbf{\bibinfo{volume}{23}}, \bibinfo{pages}{l37}
  (\bibinfo{year}{2006}{\natexlab{b}}), \eprint{gr-qc/0604018}.

\bibitem[{\citenamefont{Mishra et~al.}(2010)\citenamefont{Mishra, Arun, Iyer,
  and Sathyaprakash}}]{Mishra:2010tp}
\bibinfo{author}{\bibfnamefont{C.~K.} \bibnamefont{Mishra}},
  \bibinfo{author}{\bibfnamefont{K.~G.} \bibnamefont{Arun}},
  \bibinfo{author}{\bibfnamefont{B.~R.} \bibnamefont{Iyer}}, \bibnamefont{and}
  \bibinfo{author}{\bibfnamefont{B.~S.} \bibnamefont{Sathyaprakash}}
  (\bibinfo{year}{2010}), \eprint{1005.0304}.

\bibitem[{\citenamefont{Yunes and
  Pretorius}(2009{\natexlab{b}})}]{Yunes:2009ke}
\bibinfo{author}{\bibfnamefont{N.}~\bibnamefont{Yunes}} \bibnamefont{and}
  \bibinfo{author}{\bibfnamefont{F.}~\bibnamefont{Pretorius}},
  \bibinfo{journal}{Phys. Rev.} \textbf{\bibinfo{volume}{D80}},
  \bibinfo{pages}{122003} (\bibinfo{year}{2009}{\natexlab{b}}),
  \eprint{0909.3328}.

\bibitem[{\citenamefont{Cornish et~al.}(2011)\citenamefont{Cornish, Sampson,
  Yunes, and Pretorius}}]{Cornish:2011ys}
\bibinfo{author}{\bibfnamefont{N.}~\bibnamefont{Cornish}},
  \bibinfo{author}{\bibfnamefont{L.}~\bibnamefont{Sampson}},
  \bibinfo{author}{\bibfnamefont{N.}~\bibnamefont{Yunes}}, \bibnamefont{and}
  \bibinfo{author}{\bibfnamefont{F.}~\bibnamefont{Pretorius}},
  \bibinfo{journal}{Phys.Rev.} \textbf{\bibinfo{volume}{D84}},
  \bibinfo{pages}{062003} (\bibinfo{year}{2011}), \eprint{1105.2088}.

\bibitem[{\citenamefont{Li et~al.}(2012)\citenamefont{Li, Del~Pozzo, Vitale,
  Van Den~Broeck, Agathos et~al.}}]{Li:2011cg}
\bibinfo{author}{\bibfnamefont{T.}~\bibnamefont{Li}},
  \bibinfo{author}{\bibfnamefont{W.}~\bibnamefont{Del~Pozzo}},
  \bibinfo{author}{\bibfnamefont{S.}~\bibnamefont{Vitale}},
  \bibinfo{author}{\bibfnamefont{C.}~\bibnamefont{Van Den~Broeck}},
  \bibinfo{author}{\bibfnamefont{M.}~\bibnamefont{Agathos}},
  \bibnamefont{et~al.}, \bibinfo{journal}{Phys.Rev.}
  \textbf{\bibinfo{volume}{D85}}, \bibinfo{pages}{082003}
  (\bibinfo{year}{2012}), \eprint{1110.0530}.

\bibitem[{\citenamefont{Chatziioannou et~al.}(2012)\citenamefont{Chatziioannou,
  Yunes, and Cornish}}]{Chatziioannou:2012rf}
\bibinfo{author}{\bibfnamefont{K.}~\bibnamefont{Chatziioannou}},
  \bibinfo{author}{\bibfnamefont{N.}~\bibnamefont{Yunes}}, \bibnamefont{and}
  \bibinfo{author}{\bibfnamefont{N.}~\bibnamefont{Cornish}},
  \bibinfo{journal}{Phys.Rev.} \textbf{\bibinfo{volume}{D86}},
  \bibinfo{pages}{022004} (\bibinfo{year}{2012}), \eprint{1204.2585}.

\bibitem[{\citenamefont{Cornish and Littenberg}(2007)}]{cornish:083006}
\bibinfo{author}{\bibfnamefont{N.~J.} \bibnamefont{Cornish}} \bibnamefont{and}
  \bibinfo{author}{\bibfnamefont{T.~B.} \bibnamefont{Littenberg}},
  \bibinfo{journal}{Physical Review D (Particles, Fields, Gravitation, and
  Cosmology)} \textbf{\bibinfo{volume}{76}}, \bibinfo{eid}{083006}
  (pages~\bibinfo{numpages}{11}) (\bibinfo{year}{2007}),
  \urlprefix\url{http://link.aps.org/abstract/PRD/v76/e083006}.

\bibitem[{\citenamefont{{Del Pozzo} et~al.}(2011)\citenamefont{{Del Pozzo},
  {Veitch}, and {Vecchio}}}]{2011arXiv1101.1391D}
\bibinfo{author}{\bibfnamefont{W.}~\bibnamefont{{Del Pozzo}}},
  \bibinfo{author}{\bibfnamefont{J.}~\bibnamefont{{Veitch}}}, \bibnamefont{and}
  \bibinfo{author}{\bibfnamefont{A.}~\bibnamefont{{Vecchio}}},
  \bibinfo{journal}{ArXiv e-prints}  (\bibinfo{year}{2011}),
  \eprint{1101.1391}.

\bibitem[{\citenamefont{Dickey}(1971)}]{Dickey}
\bibinfo{author}{\bibfnamefont{J.~M.} \bibnamefont{Dickey}},
  \bibinfo{journal}{Ann. Math. Statist.} \textbf{\bibinfo{volume}{42}},
  \bibinfo{pages}{204} (\bibinfo{year}{1971}).

\bibitem[{\citenamefont{Sambridge et~al.}(2006)\citenamefont{Sambridge,
  Gallagher, Jackson, and Rickwood}}]{Sambridge}
\bibinfo{author}{\bibfnamefont{M.}~\bibnamefont{Sambridge}},
  \bibinfo{author}{\bibfnamefont{K.}~\bibnamefont{Gallagher}},
  \bibinfo{author}{\bibfnamefont{A.}~\bibnamefont{Jackson}}, \bibnamefont{and}
  \bibinfo{author}{\bibfnamefont{P.}~\bibnamefont{Rickwood}},
  \bibinfo{journal}{Geophysical Journal International}
  \textbf{\bibinfo{volume}{16}}, \bibinfo{pages}{528} (\bibinfo{year}{2006}).

\bibitem[{\citenamefont{{Yunes} and {Hughes}}(2010)}]{2010PhRvD82h2002Y}
\bibinfo{author}{\bibfnamefont{N.}~\bibnamefont{{Yunes}}} \bibnamefont{and}
  \bibinfo{author}{\bibfnamefont{S.~A.} \bibnamefont{{Hughes}}},
  \bibinfo{journal}{\prd} \textbf{\bibinfo{volume}{82}},
  \bibinfo{pages}{082002} (\bibinfo{year}{2010}), \eprint{1007.1995}.

\bibitem[{\citenamefont{{Damour} et~al.}(2000)\citenamefont{{Damour}, {Iyer},
  and {Sathyaprakash}}}]{2000PhRvD..62h4036D}
\bibinfo{author}{\bibfnamefont{T.}~\bibnamefont{{Damour}}},
  \bibinfo{author}{\bibfnamefont{B.~R.} \bibnamefont{{Iyer}}},
  \bibnamefont{and} \bibinfo{author}{\bibfnamefont{B.~S.}
  \bibnamefont{{Sathyaprakash}}}, \bibinfo{journal}{\prd}
  \textbf{\bibinfo{volume}{62}}, \bibinfo{eid}{084036} (\bibinfo{year}{2000}),
  \eprint{arXiv:gr-qc/0001023}.

\bibitem[{\citenamefont{Cornish}(2010)}]{Cornish:2010kf}
\bibinfo{author}{\bibfnamefont{N.~J.} \bibnamefont{Cornish}}
  (\bibinfo{year}{2010}), \eprint{1007.4820}.

\bibitem[{\citenamefont{Vallisneri}(2011)}]{Vallisneri:2011ts}
\bibinfo{author}{\bibfnamefont{M.}~\bibnamefont{Vallisneri}},
  \bibinfo{journal}{Phys.Rev.Lett.} \textbf{\bibinfo{volume}{107}},
  \bibinfo{pages}{191104} (\bibinfo{year}{2011}), \eprint{1108.1158}.

\bibitem[{\citenamefont{Nissanke et~al.}(2010)\citenamefont{Nissanke, Holz,
  Hughes, Dalal, and Sievers}}]{Nissanke:2009kt}
\bibinfo{author}{\bibfnamefont{S.}~\bibnamefont{Nissanke}},
  \bibinfo{author}{\bibfnamefont{D.~E.} \bibnamefont{Holz}},
  \bibinfo{author}{\bibfnamefont{S.~A.} \bibnamefont{Hughes}},
  \bibinfo{author}{\bibfnamefont{N.}~\bibnamefont{Dalal}}, \bibnamefont{and}
  \bibinfo{author}{\bibfnamefont{J.~L.} \bibnamefont{Sievers}},
  \bibinfo{journal}{Astrophys.J.} \textbf{\bibinfo{volume}{725}},
  \bibinfo{pages}{496} (\bibinfo{year}{2010}), \eprint{0904.1017}.

\end{thebibliography}
\end{document}